# A Comment on "Absence of detectable current-induced magneto-optical Kerr effects in Pt, Ta and W" by Riego *et al*

O.M.J. van 't Erve, A.T. Hanbicki, K.M. McCreary, C.H. Li and B.T. Jonker
Materials Science & Technology Division
*Naval Research Laboratory*, Washington, DC 20375  USA

We recently reported measurements of spin polarization in $\beta$-W and Pt thin films produced by the spin Hall effect (SHE) using a magneto-optic Kerr effect (MOKE) system based on crossed polarizers that detects changes in light intensity.[1]  Riego *et al* used a generalized magneto-optical ellipsometry system that in principle can distinguish pure optical reflectivity from magneto-optic signals, but were unable to detect SHE polarization in their nominally $\beta$-W, $\beta$-Ta and Pt films.[2]  They argued that our results are spurious and likely due to resistive heating which temporally modulates the film temperature and reflectivity, and that any SHE polarization is too small to be detected in metal films.

In this comment, we argue that our original results are correct as presented, and discuss why Riego *et al* were unable to detect such polarization.  First, we showed that our MOKE signal was maximum when the light propagation direction and SHE spin polarization were parallel, and nearly zero when they were perpendicular, as expected.  Local heating cannot explain such data. Second, we used a 0.5 Hz bipolar square wave for the bias current, so that any resistive heating is constant and the sample is at a constant temperature.  Third, we reported a SHE MOKE signal from a Pt film known to have a significant spin Hall angle, but could detect no signal from $\alpha$-W or Al reference films with similar resistivities, where any spurious heating effects should be comparable.  Fourth, we demonstrated a linear dependence of the MOKE signal on the magnitude of the SHE bias current, which Riego *et al* argued was characteristic of a true magneto-optic rather than reflectivity signal.  Fifth, we have confirmed our results by repeating our measurements using a commercial MOKE system[3] that uses a spatial light modulator to distinguish Kerr rotation from spurious changes in reflected intensity.  The results are shown in Figure 1a for a $\beta$-W film with bipolar square wave bias current.  When the incident laser light is parallel (blue) to the SHE polarization, the MOKE signal clearly tracks the bias current.  When the incident laser light is perpendicular (red) to the polarization, the MOKE output shows no correlation with the bias current, and only noise is observed. Sixth, we show that the effect of temperature variation is distinctly different in character by using a unipolar square wave bias current to purposefully modulate the film temperature. Figure 1b shows that on the 1 sec time scale of interest, the detected signal exhibits a linear rather than square wave behavior, does not saturate, can be an order of magnitude larger, and is independent of the angle between the incident laser light and the bias current. Although the instrument is not immune to such contributions, they can be readily distinguished from true polarization contributions.

We believe there are several potential reasons why Riego *et al* could not detect SHE polarization in their films. First, they did not report any experimentally measured non-zero SHE MOKE signal to confirm the sensitivity of their apparatus. Second, they studied only one W film thickness of 15 nm. We have found that the SHE MOKE signal is highly sensitive to the W film resistivity and thickness due to the metastable nature of β-W, as shown in Figure 2. For our film synthesis conditions, the signal decreases very rapidly for thicknesses greater than 8 nm, attributed to an increasing film fraction of the α-W phase. Third, local heating caused by their laser spot may have converted any β-W to α-W which exhibits no SHE polarization. If their laser spot was small compared to the sample size, this would not have been detected as a change in resistivity. Finally, we believe that they miscalculated the expected magnitude of SHE polarization in thin metal films. They argue that only a small fraction of the available electrons participate in SHE. However, SHE will polarize every available free electron. The spin accumulation is given by $\Delta\mu = 2\vartheta_{SH}\lambda_S j_c \rho$.[4] In β-W with a spin Hall angle $\vartheta_{SH}$ = 0.3, spin diffusion length $\lambda_S$ = 10 nm, resistivity $\rho$ = 200 $\mu\Omega$cm and charge current density $j_c$ = 1.5 × $10^6$ A/cm$^2$, one finds $\Delta\mu$ = 180 $\mu$eV, or a spin density of 1.8 x $10^{19}$ cm$^3$ for the bulk. This is 0.02% or 3 orders of magnitude smaller than the spin polarization in a ferromagnet, but well within the capabilities of either home-made or commercial MOKE systems.

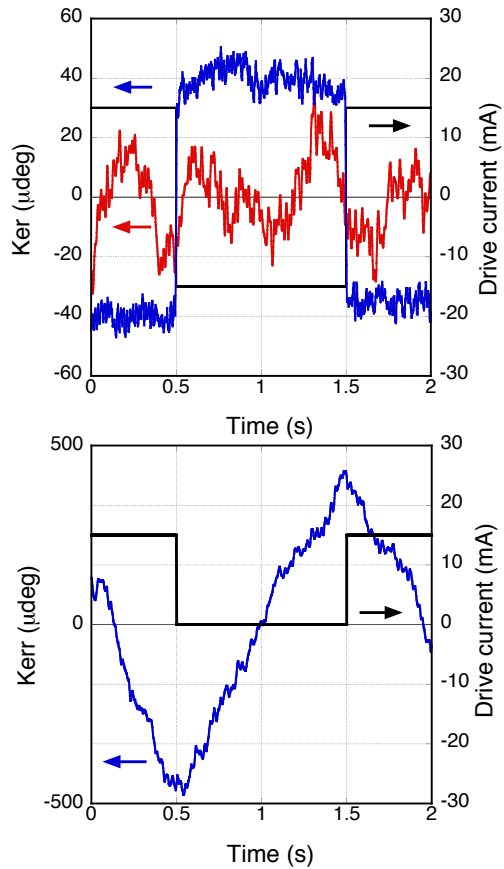

**Figure 1.** Room temperature Kerr signal *vs* time for an 8 nm β-W film. a) Bipolar drive current (black) and MOKE data for current perpendicular to plane of incidence (blue) so that SHE polarization is parallel to laser propagation direction as in longitudinal MOKE, and for current parallel to plane of incidence (red). b) Data for a unipolar drive current (black) perpendicular to the plane of incidence (blue).

**Figure 2.** Longitudinal MOKE signal per unit bias current *vs* resistivity of the tungsten film. The signal peaks for a resistivity of ~ 200 µΩcm at a thickness of 8 nm for our deposition conditions. Films to the right of the maximum are thinner than the 8 nm film used in this study, and films to the left are generally thicker.

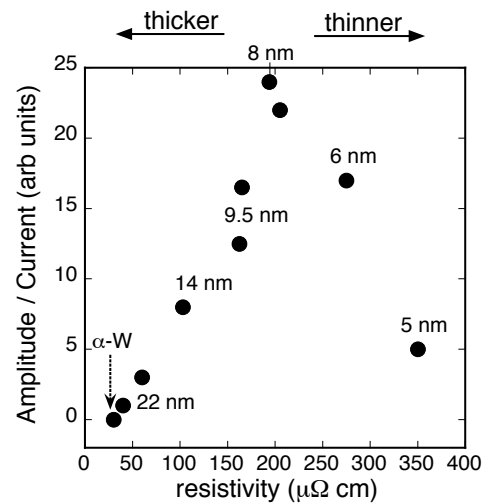